\documentclass{article}
\usepackage{cleveref}
\usepackage{natbib}
\usepackage{hyperref}

\usepackage{bm}

\begin{document}

 \begin{center}{
\bf \large Cascade processes in rapidly rotating turbulence}
\end{center}

 \begin{center}{M.Yu.~Reshetnyak \& O.~Pokhotelov  \\ 
          \vskip 0.25cm \it 
 
          Institute of Physics of the Earth RAS \\
  Moscow, Russia, m.reshetnyak@gmail.com}
\end{center}

\begin{abstract}
The process of the kinetic energy and kinetic helicity transfer over the
spectrum in an incompressible, rapidly rotating turbulent flow is
considered. An analogue of the Fjortoft theorem for 3D rapidly rotating
turbulence is proposed. It is shown that, similar to 2D turbulence, there
are two cascades simultaneously: the inverse cascade of the kinetic energy
and the direct cascade of the kinetic helicity, which in the case of 2D
turbulence corresponds to the cascade of enstrophy.
The proposed scenario is in agreement with our earlier calculations, some recent numerical simulations, and physical experiment on rotating turbulence.
 \end{abstract}

 %\copyrightstatement{TEXT}

\section{Introduction}  
The study of the spectral properties of the hydrodynamic turbulence includes
both an analysis of the kinetic energy spectra itself and the physical
fields that cause these motions, as well as the cascade processes, i.e.
transfer of certain physical quantities in the wave space. The latter, as a
     rule, reduces to study of the triad mechanisms \cite{Frisch, Verma},
arising due to interactions of the Fourier harmonics in terms, containing
products of the physical fields. In particular, for the Navier-Stokes
equation, this is a non-linear term that ensures transfer of the quantities,
that depend on the velocity field, over the spectrum. However the choice of
these quantities is quite arbitrary, usually, transfer of the integrals of
motions over the spectrum is considered. In its turn, the particular form of
the integrals of motion depend on the dimensionality of the physical space,
e.g. \cite{Les}. For 2D case they are the kinetic energy and enstrophy,
while for 3D they are  the kinetic energy and helicity.

The integral of motions are of the great importance because they determine
the direction of transfer over the spectrum, caused by the triad mechanism.
So, it is known that for 3D energy is transferred from the large scales,
where the external force is applied, to the small scales, where dissipation
is essential. The intermediate range of scales, where, in the absence of the
external forces, there is only energy transfer across the spectrum, is
called the inertial interval. This scenario of the energy transfer
corresponds to the so-called Kolmogorov's turbulence, according to which two
large vortices interacting, produce a slightly smaller vortex. Since the
scales of the vortices are close, it is customary to speak of a local energy
transfer across the spectrum.

For 2D turbulence situation is reversed: the kinetic energy is transferred
from the scale of the external force to a larger scale (the inverse energy
cascade), where there is a sink of energy that depends on the model, e.g.,
the Rayleigh friction, generation of the large-scale magnetic field in a
dynamo and etc. The presence of the inverse cascade of the energy is well
known in the atmospheric physics, where 2D approximation due to the thinness
of the atmosphere along the vertical is frequently used \cite{Pedl}.

Historically, there is an important difference in study of  2D and 3D
turbulences. For 2D, the presence of the inverse energy cascade is a
consequence of the energy conservation law and enstrophy (Fiorthof's
 theorem), e.g., \cite{Fjort}, see also \cite{Les, Ditl} and discussions
of various applications in \cite{RS, KM, Tabel}. On the contrary, the
studies of 3D turbulence were initially carried out without discussing of
kinetic helicity. It was silently assumed that in the isotropy
approximation, used by default, the mean helicity, associated with the break
of the mirror symmetry, is zero. In this case, for the non-rotating
turbulence physical and numerical experiments indicated presence of the
direct energy cascade. These results were assumed to be the general rule for
3D turbulence. The justification either rebuttal of this statement was
impossible, because  the single conservation law for kinetic energy was used.

However, in the case of the geophysical turbulence, and especially its
         extreme form in the liquid cores of the planets, where the Rossby numbers
are much less than unity, the net kinetic helicity is already non-zero.
There, in the presence of the integral of motions for 3D convection,
dependence of the fields on the coordinate along the axis of rotation
degenerates: the velocity field remains three-dimensional, but the velocity
gradients along perpendicular coordinates, with respect to the rotation
axis, are much greater than along the axis. It turns out that such an
intermediate state between 2D and 3D inherits properties of the both
two-dimensional and three-dimensional turbulence. This statement has the
direct  confirmations. Thus,  the kinetic energy fluxes in the wave
space demonstrate presence of two fluxes simultaneously: the both direct and
inverse cascades, see the results of calculations of geostrophic turbulence
  in the Boussinesq approximation in \cite{RH1, HR1} in the plane layer, and in the spherical shell \cite{ReshFScr2012}, and references their
 in.   The physical experiment with prescribed force also confirms existence of the inverse energy cascade \cite{campagne2014direct}.  Since the regimes considered above correspond to the case with a rapid
rotation with a net kinetic helicity, it is reasonable  to relate the
existence of the inverse cascade of the kinetic energy to the presence of a
second integral of motion, the conservation of the kinetic helicity. In what
follows, for the fast rotating turbulence, where velocity and vorticity
fields are strongly correlated, we demonstrate existence of two cascades: an
inverse cascade of the kinetic energy and the direct cascade of the kinetic
helicity. This approach allows to explain the earlier results of the
numerical simulations \cite{RH1, HR1}, and can be useful to analysis of
the structure of triads in the wave space.\bigskip

\section{Conservation laws and cascades}

Let us recall how the direction of the energy and enstrophy cascades in 2D
 turbulence is predicted, see for details  \cite{Les}, and
   \cite{Ditl}. It is assumed that the turbulence is homogeneous
and isotropic, so that all the fields depend on one scalar wave number $k$.
The second assumption is that only three Fourier modes with the wave numbers
$k_{1}\leq k_{2}\leq k_{3}$ interact. Then the conservation law of the
kinetic energy $E=V^{2}/2$ takes the form:
\begin{equation}
\delta E_{1}+\delta E_{2}+\delta E_{3}=0,  \label{a1}
\end{equation}%
where $\delta E_{i}$ stands for the energy fluctuations with $i=1\dots 3$.

Conservation of enstrophy $\Omega =\omega ^{2}$, where ${\bm\omega }=\nabla
\times \vec{V}$ is a vorticity, leads to the following equation for
fluctuations $\delta \Omega _{i}=k_{i}^{2}\delta E_{i}$:
\begin{equation}
\delta \Omega_{1}+\delta \Omega_{2}+\delta \Omega_{3}=0.  \label{a2}
\end{equation}%
For the particular wave numbers $k_{2}=2k_{1}$ and $k_{3}=3k_{1}$,
fluctuations $\delta E$, and $\delta \Omega $ are related as:
\begin{equation}
\begin{array}{l}
\displaystyle\delta E_{1}=-{\frac{5}{8}}\delta E_{2},\qquad \delta E_{3}=-{%
\frac{3}{8}}\delta E_{2}, \\
\\
\displaystyle\delta \Omega _{1}=-{\frac{5}{32}}\delta \Omega _{2},\qquad
\delta \Omega _{3}=-{\frac{27}{32}}\delta \Omega _{2}.%
\end{array}
\label{a3}
\end{equation}

Let the mode with $k_{2}$ gives away energy $\delta E_{2}<0$, and find how
it is redistributed among the modes with $k_{1}$ and $k_{3}$. Then $\delta
\Omega _{2}<0$\footnote{%
In more detail with arguments why the negative value of $\delta E_{2}$ is
chosen is considered in \cite{Les}. Briefly, it is concerned with the
irreversibility of the diffusion process of the wave packet, initially
localized at $k_{2}$ in the case of the free decaying turbulence.} and $%
\delta E_{i},\,\delta \Omega _{i}$ for $i=1$, and $i=3$ are positive. The
ratio $\delta E_{1}/\delta E_{3}=5/3>1$ corresponds to the inverse cascade
of the kinetic energy, and $\delta \Omega _{1}/\delta \Omega _{3}=5/27<1$ to
the direct cascade of the enstrophy. Below we consider how this approach,
called the Fjortoft theorem, can be adopted to 3D geostrophic turbulence,
where the conservation law for the enstrophy is replaced by the similar
equation for the kinetic helicity.

The conservation of the kinetic helicity $H=\vec{V\cdot \bm\omega }$ for
the triad interaction has the form:
\begin{equation}
\delta H_{1}+\delta H_{2}+\delta H_{3}=0,  \label{a4}
\end{equation}%
with $\delta H$ for the helicity fluctuation.

The pseudoscalar kinetic helicity is the more complex quantity than the
energy and enstrophy. Firstly, helicity in the general case, can change the
sign. But this complexity is not crucial, since it is possible to consider
domains, where $H$ possesses  a fixed sign. We assume that the sign of $H$
coincides with the sign of the net helicity over the specific volume, e.g.,
the certain hemisphere for the convection in a spherical shell, or the
half-volume in the plane layer with rigid horizontal boundaries. The accuracy of this approximation, which
corresponds to the strongly correlated velocity and vorticity fields, is
controlled in the models in the different ways. For the convection problem
with the prescribed force it is determined by the specific form of the
force. Fortunately, for this aim, it is more natural to use the
self-consistent solution to the problem of the thermal convection in the
geostrophic approximation, for which the degree of correlation $\vec{V}$
and $\bm\omega $ is extremely high \cite{HR2}. In this case $H/E\sim 1/l$,
where $l$ stands for the velocity scale, or $H\sim Ek$, where $k$ is the
wave number.

The situation is more complicated with the fundamental possibility of the
isotropy approximation application to the rapidly rotating turbulence, where
the only one scalar $k$ is used. This can be commented as follows: the
isotropy approximation for the case of the rapid rotation in some cases is
still very successful \cite{Zhou}, when we are interesting in the estimates
of the integral spectrum of the kinetic energy. Moreover, there are
indications that products of the velocity and vorticity in the rapidly
rotating flow for the different spatial directions have similar amplitudes
and spatial profiles \cite{R17}, which indirectly speaks in favor of the
scalar $k$. Further, we also use the isotropy approximation, assuming that
the vector fields depend on one scalar $k$.

Now rewrite (\ref{a4}), having in mind the particular form $\delta H$, and
that $k$ can change the sign. The latter leads to four cases $A$, $B$, $C$,
and $D$, which degenerated in 2D due to the quadratic dependence of
enstrophy on $k$:
\begin{equation}
\displaystyle p_{1}\delta E_{1}k_{1}+\delta E_{2}k_{2}+p_{2}\delta
E_{3}k_{3}=0,  \label{a5}
\end{equation}%
where pairs of constants ($p_{1},\,p_{2}$) take the values: (A) -- ($+1,\,+1$%
), (B) -- ($-1,\,+1$), (C) -- ($1,\,-1$), (D) -- ($-1,\,-1$).

Solving the system (\ref{a1}) and (\ref{a5}) with respect to $\delta E_{1}$,
and $\delta E_{3}$ with $k_{2}=2k_{1},\,k_{3}=3k_{1}$, and having in mind
relation between the energy and helicity, results in $\displaystyle{\delta
\vec{E}}=\Big(\delta E_{1},\,\delta E_{2},\,\delta E_{3}\Big)$, and $%
\displaystyle{\delta \vec{H}}=\Big(\delta H_{1},\,\delta H_{2},\,\delta
H_{3}\Big)$:
\begin{equation}
\begin{array}{l}
\displaystyle\displaystyle{\delta \vec{E}}=\left( {\frac{3p_{2}-2}{%
p_{1}-3p_{2}}},\,1,\,{\frac{2-p_{1}}{p_{1}-3p_{2}}}\right) \delta E_{2}, \\
\\
\displaystyle\displaystyle{\delta \vec{H}}=\left( p_{1}{\frac{3p_{2}-2}{%
p_{1}-3p_{2}}},\,2,\,3p_{2}{\frac{2-p_{1}}{p_{1}-3p_{2}}}\right) \delta
E_{2}.%
\end{array}
\label{a6}
\end{equation}%
The values of the variations for the cases ($A-D$) are given in Table~1.
Taking into account that $\delta E_{2}<0$, we consider direction of the
cascades in more details.

\begin{table*}[t]
\caption{Exchange of the energy, $\protect\delta E$, and kinetic helicity, $%
\protect\delta H$, between the modes $k_1,\, k_2,\, k_3$ in units of $%
\protect\delta E_2$; $\mathcal{I}$ -- inverse, $\mathcal{D}$ -- direct
cascades. }
\bigskip
\par
\begin{tabular}{|l|l|l|l|l|l|l|l|l|}
\hline
Quantity & \multicolumn{4}{c|}{$\delta E$} & \multicolumn{4}{c|}{$\delta H$}
\\ \hline
Mode/Cascade & $k_1$ & $k_2$ & $k_3$ & Cascade & $k_1$ & $k_2$ & $k_3$ &
Cascade \\ \hline
$A$ & $-{\frac{1}{2}}$ & $1$ & $-{\frac{1}{2}}$ & No & $-{\frac{1}{2}}$ & $2$
& $-{\frac{3}{2}}$ & $\mathcal{D}$ \\ \hline
$B$ & $-{\frac{1}{4}}$ & $1$ & $-{\frac{3}{4}}$ & $\mathcal{D}$ & ${\frac{1}{%
4}}$ & $2$ & $-{\frac{9}{4}}$ & $\mathcal{D}$ \\ \hline
$C$ & $-{\frac{5}{4}}$ & $1$ & ${\frac{1}{4}}$ & $\mathcal{I}$ & $-{\frac{5}{%
4}}$ & $2$ & $-{\frac{3}{4}}$ & $\mathcal{I}$ \\ \hline
$D$ & $-{\frac{5}{2}}$ & $1$ & ${\frac{3}{2}}$ & $\mathcal{I}$ & ${\frac{5}{2%
}}$ & $2$ & $-{\frac{9}{2}}$ & $\mathcal{D}$ \\ \hline
Sum over $A$, $B$, $C$, $D$ & $-{\frac{9}{2}}$ & $4$ & ${\frac{1}{2}}$ & $%
\mathcal{I}$ & $1$ & $8$ & $-9$ & $\mathcal{D}$ \\ \hline
\end{tabular}%
\end{table*}

For $A $, there is no cascade, since the energy from $k_2 $ is equally
distributed between $k_1 $, and $k_3 $. The helicity cascade is direct,
since $k_3 $ has got helicity three times larger than $k_1 $.

For $B $, the cascade of energy is direct, since the mode $k_3 $ received
three times more energy than $k_1 $. For helicity, the cascade is also
direct: the $k_1 $ and $k_2 $ modes transferred the helicity to the $k_3 $
mode.

For $C $, the $k_2 $ and $k_3 $ modes transmit the energy to the $k_1 $
mode, the inverse cascade of energy is accompanied by the inverse helicity
cascade. For $D $, inverse and direct cascades of energy and helicity are
observed, respectively.

Since $\delta E_2 $ is the same in $A-D $, we find the total variations,
assuming that all the cases have the same probability, see Table~1. For the
energy, the total cascade turns out to be inverse, and for helicity it is direct.
The situation resembles the case for 2D, but instead of the inverse
enstrophy cascade there is a helicity cascade.

To estimate how far the results depend on the specific choice of the values
of $k_{i}$ in triads, we rewrite (\ref{a5})  as follows:
\begin{equation}
p_{1}\alpha \delta E_{1}+\beta \delta E_{2}+p_{2}\delta E_{3}=0,  \label{a7}
\end{equation}%
where real $\alpha $, and $\beta $ satisfy to the conditions: $0\leq \beta
\leq 1$, $0\leq \alpha \leq \beta $. This approximation is valid for the
large wave numbers with $\alpha \sim k_{1}/k_{3},\,\beta \sim k_{2}/k_{3}$,
where the discreteness of the wave numbers becomes insignificant. The joint
solution of (\ref{a1}) and (\ref{a7}) has the form:
\begin{equation}
\begin{array}{l}
\displaystyle\displaystyle{\delta \vec{E}}=\left( {\frac{\beta -p_{2}}{%
p_{2}-\alpha p_{1}}},\,1,\,{\frac{\alpha p_{1}-\beta }{p_{2}-\alpha p_{1}}}%
\right) \delta E_{2}, \\
\\
\displaystyle\displaystyle{\delta \vec{H}}=\left( p_{1}\alpha {\frac{%
\beta -p_{2}}{p_{2}-\alpha p_{1}}},\,\beta ,\,p_{2}{\frac{\alpha p_{1}-\beta
}{p_{2}-\alpha p_{1}}}\right) \delta E_{2}.%
\end{array}
\label{a8}
\end{equation}%
Expressions (\ref{a8}) describe the energy and helicity fluctuations for
arbitrary relations of the triangle' sides $(\alpha ,\,\beta ,\,1)$.

Note that according to (\ref{a8}), the non-local cascade of the kinetic
energy, predicted in the more sophisticated models \cite{Waleff}, can
exist. Let $\alpha \rightarrow 0$, $\beta \rightarrow 1$, then $\displaystyle%
{\delta \vec{E}}=\left( {\frac{1-p_{2}}{p_{2}}},\,1,\,-{\frac{1}{p_{2}}}%
\right) \delta E_{2}$, and for $p_{2}=-1$ the non-local inverse cascade of
the kinetic energy takes place, where modes $k_{2}$, $k_{3}$ feed mode $k_{1}
$.

Recalling that cases $A-D$ are equiprobable, we summarize each of the
components of the vectors ${\delta \vec{E}}$ and ${\delta \vec{H}}$ in
$p_{1}$ and $p_{2}$, and divide it by 4, the number of cases:
\begin{equation}
\begin{array}{l}
\displaystyle\displaystyle{\delta \vec{E}}^{\Sigma }=\left( -{\frac{1}{%
1-\alpha ^{2}}},\,1,\,{\frac{\alpha ^{2}}{1-\alpha ^{2}}}\right) \delta
E_{2}, \\
\\
\displaystyle\displaystyle{\delta \vec{H}}^{\Sigma }=\beta \left( {\frac{%
\alpha ^{2}}{1-\alpha ^{2}}},\,1,\,-{\frac{1}{1-\alpha ^{2}}}\right) \delta
E_{2}.%
\end{array}
\label{a9}
\end{equation}%
Note that, as $0<\alpha <1$, and $\delta E_{2}<0$, the components of the
vector ${\delta E}_{2}^{\Sigma }=\delta E_{2}<0$, ${\delta E}_{3}^{\Sigma }<0
$, and $\displaystyle{\delta E}_{1}^{\Sigma }={\frac{\alpha ^{2}}{1-\alpha
^{2}}}\delta E_{2}=-({\delta E}_{2}^{\Sigma }+{\delta E}_{3}^{\Sigma })>0$.
This corresponds to the inverse cascade of the energy. For $\alpha
 \rightarrow 0$ the total cascade from $k_{3}$ is zero. So far $0<\alpha<\beta<1$, the   non-local cascade between $k_1$ and $k_2$ exists. 

For the kinetic helicity situation is quite opposite: $\displaystyle {\delta
H}^\Sigma_1= {\frac{ \alpha^2\beta }{1- \alpha^2 }} \delta E_2 <0$, ${\delta
H}^\Sigma_2=\beta \delta E_2 <0$, and $\displaystyle {\delta H}^\Sigma_3= - {%
\frac{ \beta }{1- \alpha^2 }} \delta E_2 = -({\delta H}^\Sigma_1+ {\delta H}%
^\Sigma_2 )>0$, the direct cascade of helicity takes place.

It is instructive to examine the case with exchange between $k_1$ and $k_3$,
and $k_2$ for a catalyst. For energy this corresponds to the case when $%
\alpha \to 1 $. In this case, the mode with $k_2 $ does not participate in
the energy exchange, and all energy from $k_3 $ goes to $k_1 $. Since $%
\beta> \alpha $, so that the both $\alpha\to 1$ and $\beta\to 1$, this
corresponds to an equilateral triangle in the wave space, and to the local
transfer, correspondingly.

For helicity, a similar analysis yields: for $\alpha \to 1 $, the $k_2 $
mode participates in the exchange, and for $\alpha \to 0 $ mode $k_1 $ does
not involve in the exchange.

Expressions (\ref{a9}) describe the energy and helicity fluctuations for
arbitrary relations of the sides of the triangle $(\alpha ,\,\beta ,\,1)$.
To find the total contribution for various $\alpha $, and $\beta $ one has
to integrate over $\alpha $, and $\beta $. Let introduce the functional $%
\mathcal{I}(f)=\int\limits_{0}^{1}\int\limits_{0}^{\beta }f(\alpha ,\,\beta
)\,d\alpha d\beta $, and calculate it for fluctuations $\displaystyle%
(1-\alpha ^{2}){\delta E}_{i}^{\Sigma }$, $\displaystyle(1-\alpha ^{2}){%
\delta H}_{i}^{\Sigma }$, substituting them instead of $f$ in $\mathcal{I}$,
and denoting the result as $\displaystyle{\delta \vec{E}}^{\int }$, $%
\displaystyle{\delta \vec{H}}^{\int }$, respectively:
\begin{equation}
\begin{array}{l}
\displaystyle\displaystyle{\delta \vec{E}}^{\int }=\left( -{\frac{6}{12}}%
,\,{\frac{5}{12}},\,{\frac{1}{12}}\right) \delta E_{2}, \\
\\
\displaystyle\displaystyle{\delta \vec{H}}^{\int }=\left( {\frac{1}{15}}%
,\,{\frac{4}{15}},\,-{\frac{5}{15}}\right) \delta E_{2}.%
\end{array}
\label{a10}
\end{equation}%
 Relations (\ref{a10}) give an idea of the total ratio of the contributions of
modes $k_{i}$ to the inverse cascade of kinetic energy and to the direct
cascade of the kinetic helicity over the spectrum. As it was predicted by
estimates with the particular choice of the wave numbers above, there are
two cascades of the kinetic energy and helicity, one of that is direct, and
the second one is inverse. For the both cascades all three modes participate
in the exchange. Contribution of the intermediate mode $k_2$ is comparable to the both other modes. \bigskip

 \section{Conclusions}  %% \conclusions[modified heading if necessary]
     Inspection of the recent papers  \cite{mininni2010rotating,biferale2017two} on cascades in 3D rotating turbulence reveals that inverse cascade of the kinetic energy is quite common phenomenon, however its details depend on the parameter regimes, such as, e.g., the  angular rotation velocity \cite{mininni2009scale,campagne2014direct}, scale of the prescribed force \cite{deusebio2014dimensional}. In this sense the considered above simple scenario of the inverse cascade for the energy reflects general properties of the rotating turbulence, and  seems to be quite natural. Dependence on parameters make study more complicated and one way to overcome this difficulty is to consider more realistic forms of the prescribed force. By this we mean that the force should be tested for consistency with the self-consistent natural convection, e.g. thermal convection in presence of the rapid rotation. Such test will exclude possibility when the increase of the angular rotation velocity will cause transition from 3D to 2D. This situation with unclear transition from one set of integral of motions to the other one, can be unreal in principle. Thus, for the thermal convection (the same for compositional one)   increase of rotation  suppresses convection. But if thermal convection has already started in the system with collinear  gravity and axis of rotation, then the flow is geostrophic but still three-dimensional with similar in amplitudes of velocity components, see \cite{roberts1965thermal}. Consequently, the models, which demonstrate something different, usually omit buoyancy at all, either do not take into account dependence of the critical Rayleigh number on rotation. The back side of the coin is that fully self-consistent models, as a rule, have worse spatial resolution, and may be what is more important, in such models, due to absence of the sink of the energy at large scales, the statistical equilibrium there is developed \cite{Ditl},  decreasing amplitude of the inverse flux, and  making  interval of the spectra, where the inverse cascade takes place, even shorter. The obtained  results can be interesting  for testing  of the turbulent diffusion coefficients in geostrophic flows, see for details \cite{smith2002turbulent}.

  %\bibliography{npg.bib}

\end{document}